\documentclass[pmlr]{jmlr}% new name PMLR (Proceedings of Machine Learning)

\makeatletter
\def\set@curr@file#1{\def\@curr@file{#1}} %temp workaround for 2019 latex release
\makeatother
\usepackage[load-configurations=version-1]{siunitx} % newer version
\usepackage{lipsum}
\usepackage{mwe}
\usepackage{tikz}
\makeatletter
\newcommand*{\centernot}{%
  \mathpalette\@centernot
}
\def\@centernot#1#2{%
  \mathrel{%
    \rlap{%
      \settowidth\dimen@{$\m@th#1{#2}$}%
      \kern.5\dimen@
      \settowidth\dimen@{$\m@th#1=$}%
      \kern-.5\dimen@
      $\m@th#1\not$%
    }%
    {#2}%
  }%
}
\makeatother

\usepackage{mathtools}

\DeclarePairedDelimiterX\Basics[1](){ #1}

\usepackage{algorithm}
\usepackage{algorithmic}
\usepackage{enumitem}
\usepackage{multirow}

\usepackage{longtable}% for long tables

\usepackage{booktabs}
\usepackage[load-configurations=version-1]{siunitx} % newer version
\theorembodyfont{\upshape}
\theoremheaderfont{\scshape}
\theorempostheader{:}
\theoremsep{\newline}

\jmlrvolume{XXX}
\jmlryear{2022}
\jmlrworkshop{Machine Learning for Healthcare}

\title[PCT through SCM]{Pragmatic Clinical Trials in the Rubric of \titlebreak Structural Causal Models}

\author{\Name{Riddhiman Adib} \Email{riddhiman.adib@marquette.edu} \\
      \addr Department of Computer Science\\
      Marquette University\\
      Milwaukee, Wisconsin, USA
      \AND
      \Name{Sheikh Iqbal Ahamed} \Email{sheikh.ahamed@marquette.edu} \\
      \addr Department of Computer Science\\
      Marquette University\\
      Milwaukee, Wisconsin, USA
      \AND
      \Name{Mohammad Adibuzzaman} \Email{adibuzza@ohsu.edu} \\
      \addr Oregon Clinical and Translational Research Institute\\
      Oregon Health \& Science University\\
      Portland, Oregon, USA
}

\begin{document}

\maketitle

% Tells us a bit about the problem.  Recent advances in machine learning
% \citep{cite1} have resulted in great things happening in healthcare.
% In particular, \citet{cite2} describes a spiffy technique to save even
% more lives.  In this work, we...

% ARXIV VERSION
\begin{abstract}
    Explanatory studies, such as randomized controlled trials, are targeted to extract the true causal effect of interventions on outcomes and are by design adjusted for covariates through randomization. On the contrary, observational studies are a representation of events that occurred without intervention. Both can be illustrated using the Structural Causal Model (SCM), and do-calculus can be employed to estimate the causal effects. Pragmatic clinical trials (PCT) fall between these two ends of the trial design spectra and are thus hard to define. Due to its pragmatic nature, no standardized representation of PCT through SCM has been yet established. In this paper, we approach this problem by proposing a generalized representation of PCT under the rubric of structural causal models (SCM). We discuss different analysis techniques commonly employed in PCT using the proposed graphical model, such as intention-to-treat, as-treated, and per-protocol analysis. To show the application of our proposed approach, we leverage an experimental dataset from a pragmatic clinical trial. Our proposition of SCM through PCT creates a pathway to leveraging do-calculus and related mathematical operations on clinical datasets.
\end{abstract}

\section{Introduction}

% RCT, PCT, and OBS
Experimental studies with varying designs and research goals, such as Randomized controlled trials (RCT), Pragmatic clinical trials (PCT), are frequently conducted in many branches of science to derive the causal effect of interventions. Conversely, observational studies (OBS) capture the outcome of an incident without any alteration of the independent variable. Due to differences in the experiment settings (e.g., goal, population group, treatment protocol), the causal findings of the experiments are harder to compare, and the transfer of knowledge from one study population to another is not very trivial. Thus, there is a need for generalizability or structural methodology to draw unbiased causal inferences from experiments (RCT+PCT+OBS), leveraging their unique design attributes. 

% Causal effect estimation and SCM
In recent times, through the advancement of machine learning and artificial intelligence, finding newer ways of causal explorations from datasets available, i.e., data-driven causal inference, is of high interest. Structural Theory of Causation (SCM), proposed by Judea Pearl \citep{pearl2016causal} and extended by many other researchers \citep{bareinboim2016causal}, holds the potential to define scientific studies for causal inference, express them through graphical causal models and transfer knowledge in between them.

% Why our work - what are we doing - Need for it - Why PCT differs from RCT and OBS
SCMs allow researchers to represent scientific studies systematically. SCM representation of an experimental study helps portray underlying causal mechanisms, express causal effects of interventions and answer hypothetical questions. However, core differences of PCT from RCT and OBS \citep{gamerman2019pragmatic}, in terms of (a) population, (b) setting, (c) comparison arm (treatment), and (d) outcome, makes it challenging for objective evaluation of interventions and their effect. This paper illustrates a causal representation of PCT and relevant mathematical formulations to aid causal effect estimations and objective evaluations in a target population and interpret existing analysis techniques through a causal lens.

% current state
Contrary to RCT and OBS frequently being formulated through SCM \citep{pearl2016causal}, representation of PCT with SCM is still an ongoing research problem \citep{hernan2017per}. Additionally, novel ways of utilizing priors (background knowledge) to build a comprehensive causal model from data is also under exploration \citep{nordon2019building}. In summary, a standardized way to represent PCT through SCM is not yet fully grounded on the theories of recent advances in causal inference.

%=====================================================
\section{Background}~\label{sec:background}

This section describes the relevant background concepts, such as various scientific studies, including pragmatic clinical trials and their unique attributes. We then discuss the structural theory of causal within causal inference and structural causal models for various scientific studies. Finally, we present our problem formulation, followed by related works.

\subsection{Experimental Studies}
% definition
In broader terms, based on design factors, scientific studies follow two routines: experimental studies and observational studies. Experimental studies are at the core of most scientific investigations. In experimental studies, experimenters introduce a dependent variable (e.g., treatment or procedure) and consecutively observe an outcome \citep{coggon2009epidemiology}. Most commonly, the underlying research question is uncovering the effect of an outcome compared to an intervention or factor. 

% types of study and diff than OBS
The design of experimental studies is a well-explored research area \citep{fisher1936design,eriksson2000design}. The most popular and effective experimental study, especially to find causal relations, is the randomized controlled trial (RCT) \citep{grossman2005randomized}. In RCTs, researchers explore the effects of treatment on outcome in a narrower population (with clear and strict inclusion-exclusion criteria) with randomization (to control for both known and unknown confounding) \citep{gamerman2019pragmatic}. RCTs are harder to implement and cost more; however, they unquestionably justify the causal effect of treatment by comparing treatment arms.

% OBS
Since experimental studies require significant resources (in time and expenses) and are sometimes unethical or infeasible for a certain population, researchers occasionally conduct studies through exploring existing real-world data (e.g., EHR dataset) collected without any intervention. These studies are called observational studies (OBS) (or, natural experiments) \citep{rosenbaum2005observational}. RCT and OBS are inherently different from each other, the two prime differences being (a) presence of intervention, (b) de-confounding through randomization. In general, RCTs are considered as a higher level of evidence compared to OBS \citep{concato2004observational}.

% pct is diff than rct
One other type of experimental study is pragmatic clinical trials (PCT). By nature, PCTs are more fluid and have characteristics floating between an RCT and an OBS. 

\subsection{Pragmatic Clinical Trials}

\subsubsection{Definition}
Pragmatic clinical trials (PCT) are a variety of experimental studies that aim to explore correlations between treatment and outcome in a real-world health system, contrary to focusing on causal explorations \citep{murray2019guidelines}. Uncovering causal effects through experimental studies requires extreme deconfounding and strict inclusion-exclusion criteria, sometimes making the study result irrelevant to real-world practice. The goal is to define clinical decision-making rather than regulatory approval. Two significant challenges of PCT are: (1) missing data and (2) non-adherence to protocol.

\subsubsection{Features}
\begin{figure}[htbp]
    \centering
    \includegraphics[width=0.5\textwidth]{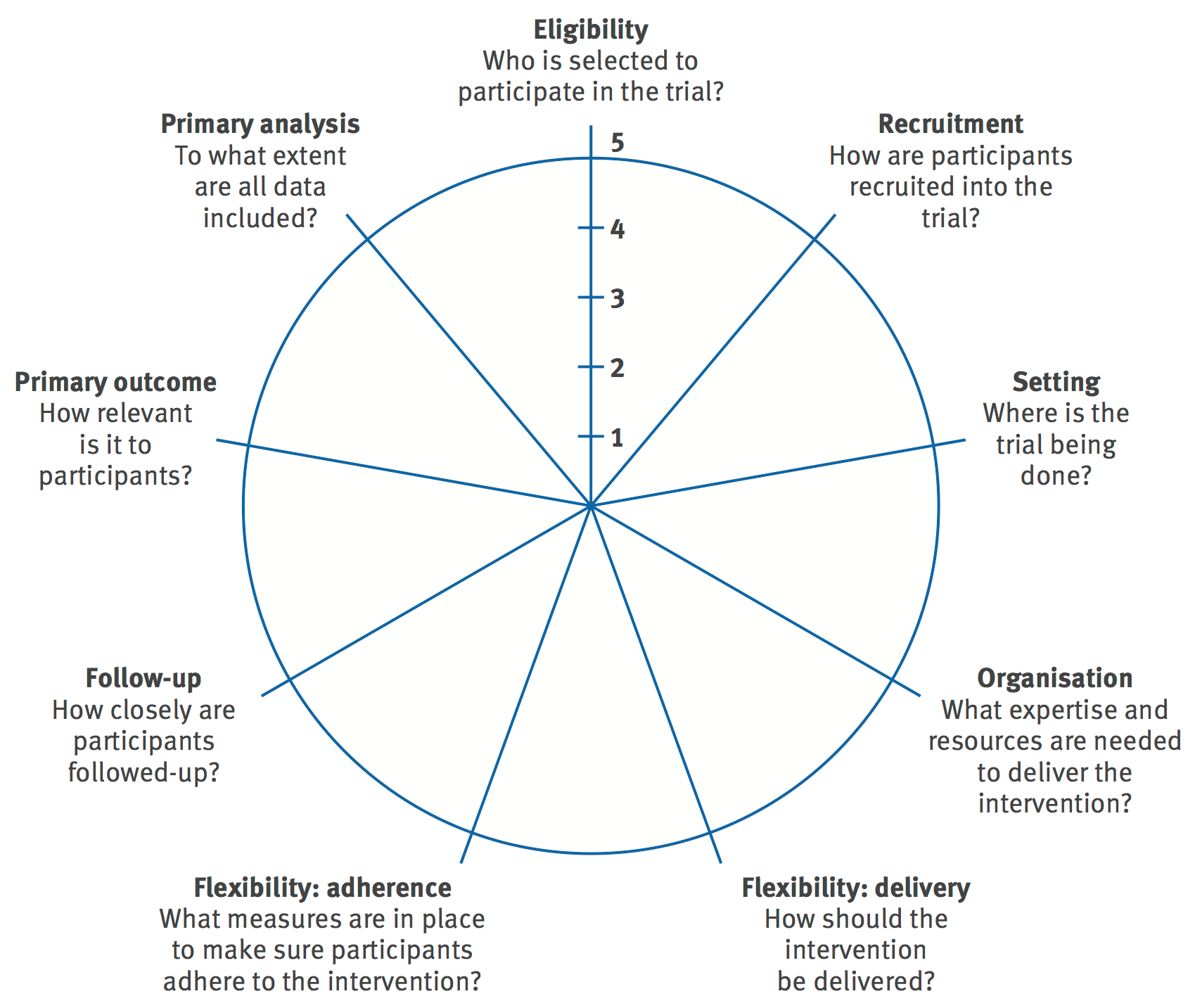}
    \caption{Visualization of PRECIS (PRagmatic Explanatory Continuum Indicator Summary)}
    ~\label{fig:precis}
\end{figure}

Due to its pragmatic nature, features of PCT have drawn much discussion from the scientific community. \citep{gamerman2019pragmatic} defines PCT as a variation of RCT, with four critical \textit{pragmatic} design elements: (a) real-world population (recruitment extended to fit all potentially eligible individuals receiving care in participating setting), (b) real-world setting (commonly takes place in a flexible setting closer to patients' usual clinical care, avoiding the need for specially trained research staff for data collection), (c) appropriate comparison arm (sometimes combining multiple drugs or multiple doses of the same drug), and (d) relevant outcome (goal is to understand the real-world implications of the intervention). \citep{loudon2015precis,purgato2015pragmatic} have laid out nine features of PCT, depicted as a wheel in \autoref{fig:precis} (lower score signifies explanatory and higher axis signifies pragmatism in nature). 

\subsubsection{Analysis Methods}
Since the treatment population group varies based on adherence and loss-to-follow-up in PCTs, various analysis protocol is followed in the data investigation of PCTs. The three most common analysis protocols for PCT are (1) Intention-to-treat (ITT), (2) As-treated (AT), and (3) Per-protocol (PP).

In Intention-to-treat (ITT) analysis, all randomized patients are included, regardless of whether they adhered to the treatment prescribed or subsequent withdrawal \citep{gupta2011intention,montori2001intention}. It essentially ignores anything after randomization (e.g., withdrawal, protocol non-compliance), and in general, avoids overoptimistic estimates of the intervention efficacy. For this reason, ITT is the most recommended method in PCTs \citep{gupta2011intention,boutis2011intention}. 

In As-treated (AT) analysis, patients are incorporated based on the treatment they received, irrespective of their randomization status \citep{smith2021interpreting}. Likewise, in per-protocol (PP) analysis, only those patients are included who genuinely adhered to the study prescribed, i.e., for whom the treatment prescribed and treatment received are same \citep{sedgwick2015intention}. PP analysis represents a `best-case' scenario in trial results since it represents patients who completed the treatment initially allocated and thus ignores protocol deviation or non-adherence. Both AT and PP analyses give a biased estimate of intervention efficacy; however, they are essential for the report since they reflect the impact of non-compliance and non-adherence.

\subsection{Structural Causal Models}
The structural theory of causation was proposed and established on the foundations of probabilistic graphical models by Judea Pearl \citep{pearl2016causal} and many other researchers \citep{bareinboim2016causal,spirtes2010introduction}. Under this theory, structural causal models (SCM) are a structured definition of a causal model, often portrayed through graphs. We present the formal description of an SCM \citep{pearl2009causality} in \autoref{def:scm}:

\begin{definition}[Structural Causal Model]
~\label{def:scm}
    A structural causal model $M$ is a 4-tuple \\$\langle U, V, f, P(u) \rangle $ where:
    \begin{enumerate}
        \item $U$ is a set of background (exogenous) variables that are determined by factors outside of the model,
        \item V is a set $\{V_1, V_2, ..., V_n\}$ of observable (endogenous) variables that are determined by variables in the model (i.e., determined by variables in $U \cup V$ ),
        \item F is a set of functions $\{f_1, f_2, ..., f_n\}$ such that each $f_i$ is a mapping from the respective domains of $U_i \cup P A_i$ to $V_i$, where $U_i \subseteq U$ and $P A_i \subseteq V \ V_i$ and the entire set $F$ forms a mapping from $U$ to $V$. In other words, each $f_i$ in $v_i \leftarrow f_i(pa_i, u_i), i = 1, ..., n$, assigns a value to $V_i$ that depends on the values of the select set of variables $(U_i \cup P A_i)$, and
        \item $P(u)$ is a probability distribution over the exogenous variables.
    \end{enumerate}
\end{definition}

Causal directed acyclic graphs (DAG) are commonly portrayed to express an SCM. In a causal DAG $G$, node $V$ represents an observed or unobserved variable, and directed edge $E$ represents the causal relationships between two nodes. With the purpose of investigating the causal effect of one variable on another, do-calculus was developed \citep{pearl2016causal}. Do-calculus is a multi-functional tool (mathematical formulation) to map the observational truth to the corresponding experimental reality by adjusting for different kinds of biases, such as confounding (if it exists). 

\subsection{SCM for Scientific Studies}
\begin{figure}[htbp]
    \centering
    \includegraphics[width=0.6\linewidth]{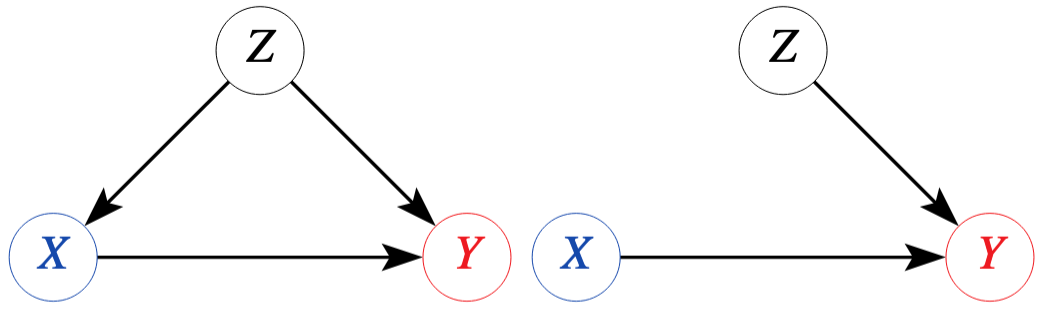}
    \caption{SCM representation of scientific studies}
    ~\label{fig:scm-obs-rct}
\end{figure}

SCM and causal DAG have been frequently used in the literature to represent various scientific studies \citep{tennant2021use}. \autoref{fig:scm-obs-rct} shows two graphical structures of SCM, one for observational study (left) and the other for randomized controlled trial (right). Both of them have treatment $X$, outcome $Y$, and confounder $Z$; the only distinction being a lack of arrow (causal connection) from $Z$ to $X$, thus representing the randomization done prior to the study. Representation through SCM helps provide a structural definition to distinct trials and allows application of do-calculus for causal effect estimation and counterfactual evaluation. 

\subsection{Problem Definition}
Since the strength of SCM in representing different studies and exploring the underlying causal mechanisms is well-established in the literature, researchers are looking for ways to represent PCT using the rubric of SCM. For this work, we focus on the two following research questions:
\begin{enumerate}
    \item how can we represent PCT through SCM?
    \item how can we represent the analysis techniques commonly deployed in PCT using SCM and do-calculus?
\end{enumerate}

\subsection{Related Works}
For causal exploration on PCT, different general guidelines have been proposed in the literature; however, a unified approach is severely lacking. \citep{hernan2017per} discussed issues involving pragmatic trials in general, along with a general causal graphical structure and adherence as a node in the graph. Without using any underlying causal structure, \citep{murray2019guidelines} have presented an elaborated guideline for a causal understanding of diverse, unique features of PCT qualitatively and figuratively to estimate the ITT and PP effects (of both point and sustained intervention). Later, in continuation to the previous two works, \citep{murray2021causal} discussed a wide variety of graphical representations of PCTs, but without employing any do-calculus for ITT or PP effect estimations. Although all the works used causal graphical structures for representing PCTs, they did not decide on a single definition or discuss its use with do-calculus for ITT, AT, or PP analysis.

%=====================================================
\section{Structural Causal Model for Pragmatic Clinical Trials}~\label{sec:method}

In this section, we introduce the notion of structural causal models (SCM) for pragmatic clinical trials (PCT). We iterate through the unique features of PCT, such as eligibility criteria, non-adherence, and loss-to-follow-up, and examine their potential interpretations in structural causal models. Following that, using the notations proposed, we discuss the frequently used analysis methods for PCT, such as intention-to-treat, as-treated, and per-protocol analysis. For simplicity, we assume a point intervention with no time-varying components (both in treatments or outcomes).

\begin{figure}[htbp]
    \centering
    \includegraphics[width=0.6\linewidth]{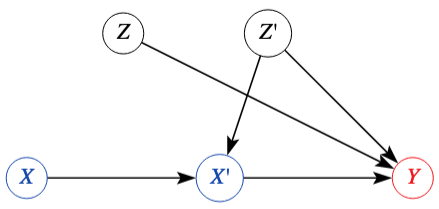}
    \caption{Graphical representation of the proposed structural causal model for pragmatic clinical trials}
    ~\label{fig:scm-pct}
\end{figure}

\subsection{Defining PCT for SCM}
To express a PCT through SCM, we start with defining the PCT. We assume, we are working with a $PCT$ with population group $\Pi$, where the query of interest is finding the effect of a treatment protocol $X$ (\textit{not the same as `causal' effect of treatment $X$, explained in \autoref{subsec:equi-rct}}) on outcome $Y$. Different arms of treatment protocol $X$ might have overlapping components, such as the same drug (or software feature) with a different dosage (or color palette). We propose that the target $PCT$ for the population $\Pi$ can be expressed through a structural causal model $M = <U, V, F, P_u>$, with a graphical representation through graph $G$, with two versions of treatment $X$ and $X'$.

\begin{figure}[htbp]
    \centering
    \includegraphics[width=0.8\linewidth]{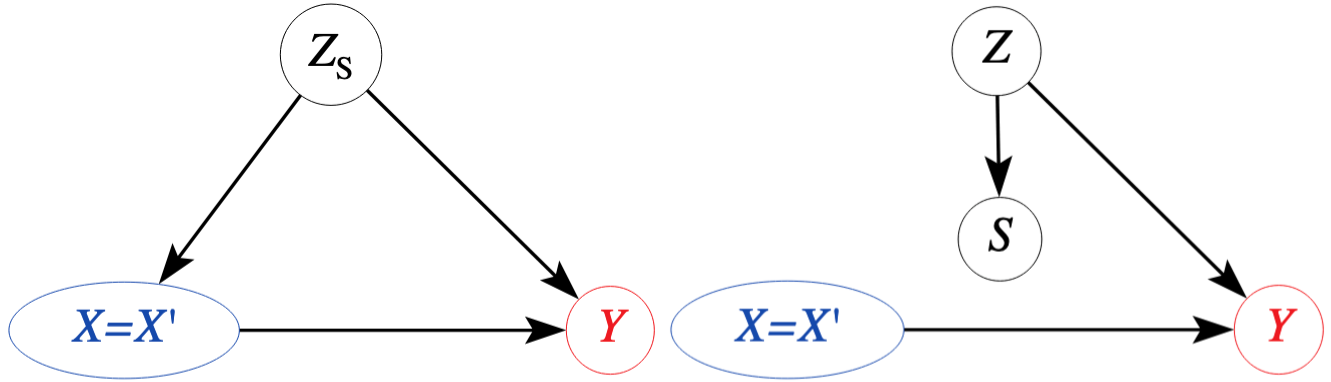}
    \caption{Graphical representation of the structural causal model of a $RCT_{PCT}$, \textit{(left)} with population $\Pi_s$ \& treatment $X$ ($= X'$), and \textit{(right)} with population $\Pi_s$ as a selection bias through node $S$ on population $Z$}
    ~\label{fig:scm-rct-pct}
\end{figure}

\subsubsection{Equivalent RCT}
~\label{subsec:equi-rct}
For reference and comparison, if the query of interest for the researchers were, in fact, finding the `causal' effect of treatment $X$ on outcome $Y$, the standard procedure would be to conduct a randomized controlled trial (RCT) on a stricter population group $\Pi_s$. In that case, the causal graph for the RCT would be similar to \autoref{fig:scm-rct-pct} (left), where treatment would be $X(=X')$. The reasoning behind having different population groups for PCT ($\Pi$) and RCT ($\Pi_s$) comes from their definitions; $\Pi_s$ would be a narrower focused group of $\Pi$ with minimal possible confounding to outcomes. We will refer to this equivalent RCT as $RCT_{PCT}$.

\subsection{Features of PCT}

\subsubsection{Treatment, Outcome and Covariates} 
A general graphical representation $G$ of the proposed SCM $M$ for PCT is presented in \autoref{fig:scm-pct}. Here, the independent variable, or treatment, is represented by $X$ \textit{(and $X'$, explained in next subsection)}, and $ Y$ represents the dependent variable or outcome. Covariates $Z$ and $Z'$ represents all other relevant variables; however, $Z$ do not have any causal effect in adherence to the trial (i.e., $X'$), whereas $Z'$ are the covariates that affect adherence (e.g., affects treatment received $X'$). 

Since $X$ is provided through randomization, there is no causal relationship ($\rightarrow$) between $Z$ or $Z'$ and $X$. However, as $Z'$ are indicators of adherence, there is a causal relationship between $Z'$ and $X'$.

\subsubsection{Non-adherence}
Since non-adherence to treatment is a core component in $PCT$, they are depicted through two separate nodes $X$ and $X'$ in the proposed causal graph. $ X$ represents the treatment prescribed (through randomization), and the treatment received (or followed by trial participants) is represented by $X'$. $X'$ is different from $X$ due to non-adherence; however, it is still influenced by $X$. The relationship between $X$ and $X'$ has previously been expressed \citep{murray2021causal} through adherence to the trial, as a percentage of adherence to the treatment prescribed.

\subsubsection{Eligibility criteria}
Compared to RCTs, PCTs are more liberal in including patients from varying demographics. As previously discussed in \autoref{sec:background}, eligibility criteria are the key reason behind this population demographic difference between a PCT and a similar RCT, and thus, between $\Pi$ and $\Pi_s$. This difference can also be viewed as a selection \citep{bareinboim2012controlling} through node $S$, where $S=1$ defines being eligible for the RCT, equivalent to the target PCT. However, selection through node $S$ in a study does not always trigger selection bias to that study.

\subsubsection{Loss-to-follow-up}
In most PCTs, population lost-to-follow-up is a concern for the scientists \citep{gamerman2019pragmatic}. Since trial participants tend to show lesser adherence to the protocol, some generally do not follow through with the treatment prescribed or disconnect with the research team and end up being the population lost-to-follow-up. During data analysis, this population data lost-to-follow-up are generally censored \citep{hernan2017per}. The conditioning of censored data can also be viewed as survivorship bias \citep{carpenter1999survivorship}, through a node $C$. For $C=1$, we select a population group who completed the trial and were not lost-to-follow-up, thus looking at a population who 'survived' the study.

\subsection{Outcome Analysis for PCT}
% Below, we discuss three standard analysis procedures employed in PCTs. 

\subsubsection{Query of Interest}
% general query
By definition, the query of interest in a PCT is finding the `effectiveness' of a treatment protocol, not the `efficacy' of specific treatment \citep{porzsolt2015efficacy}. Based on that, \citep{murray2021causal} have described the vital causal interests in a $PCT$: intention-to-treat effect, the per-protocol effect of continuous adherence to treatment versus placebo, and in general, the effect of good adherence to trial protocol versus poor adherence in the placebo arm.

\subsubsection{Intention-to-treat Analysis}
In the intention-to-treat analysis, we explore the effects on outcomes based on randomized or prescribed treatment. Since all the participants (in some cases, excluding loss-to-follow-up) are included in this analysis, we express the concern by:
\begin{equation}
~\label{eq:itt}
    P(Y | X)
\end{equation}
% How to include: $$P(Y | \mathit{do}(X))$$

\subsubsection{As-treated Analysis}
For as-treated analysis, by definition, we look for participants who indeed took the treatment rather than prescribed, so the ``true" treatment intervention would be $X'$, not $X$. For that, we express the concern by: 
\begin{equation}
~\label{eq:at}
    P(Y | X')
\end{equation}
% How to include $$P(Y | \mathit{do}(X'))$$

\subsubsection{Per-protocol Analysis}
Finally, we include the population who followed through treatment prescribed for per-protocol analysis. We exclude the population who have taken a different treatment than what was prescribed; that is, for whom $X$ and $X'$ did not match. With this, we express the concern by: 
\begin{equation}
~\label{eq:pp}
    P(Y | X=a, X'=a)
\end{equation}
% How to include $$P(Y | \mathit{do}(X=a), \mathit{do}(X'=a))$$

\subsubsection{Additional Study Metrics}
Pragmatic clinical trials additionally report other relevant study metrics, such as Odds Ratio (OR), Risk Ratio (RR), and Hazard Ratio (HR) \citep{cummings2009relative}. These metrics are used to detect the association of treatment with the outcome and provide additional insight into treatment effects. We present equations to calculate their values based on conditional probability below. In \autoref{eq:hr}, $h(t, \mathbf{X}_{x=a})$ represents hazard function with time $t$ and the vector with the covariates of the model $X$ with the value $a$ ($X = [z, x]$, where $x$ is the treatment and $z$ are the confounders).

\begin{equation}
~\label{eq:or}
    OR = \frac{\frac{P(Y=0|X=1)}{P(Y=1|X=1)}}{\frac{P(Y=0|X=0)}{P(Y=1|X=0)}}
\end{equation}

\begin{equation}
~\label{eq:rr}
    RR = \frac{\frac{P(Y=0|X=1)}{P(Y=0|X=1)+P(Y=1|X=1)}}{\frac{P(Y=0|X=0)}{P(Y=0|X=0))+P(Y=1|X=0)}}
\end{equation}

\begin{equation}
~\label{eq:hr}
    HR = \frac{h(t, \mathbf{X}_{x=1})}{h(t, \mathbf{X}_{x=0})}
\end{equation}

% \begin{equation}
% ~\label{eq:cor}
%     COR = \frac{\frac{P(Y=0|X=1))}{P(Y=1|X=1))}}{\frac{P(Y=0|X=0))}{P(Y=1|X=0))}}
% \end{equation}

% \begin{equation}
% ~\label{eq:crr}
%     CRR = \frac{\frac{P(Y=0|X=1))}{P(Y=0|X=1)+P(Y=1|X=1))}}{\frac{P(Y=0|X=0))}{P(Y=0|X=0))+P(Y=1|X=0))}}
% \end{equation}

% \begin{equation}
% ~\label{eq:chr}
%     CHR = \frac{h(t, \mathit{do}(\mathbf{X}_{x=1}))}{h(t, \mathit{do}(\mathbf{X}_{x=0}))}
% \end{equation}

Given a known structural causal model, interpreting their causal equivalent is established in the literature: causal odds ratio \& causal risk ratio \citep{palmer2011instrumental} and causally formulated hazard ratio \citep{adib2020causally}.

%=====================================================
\section{Example of PCT with SCM}

In this section, we apply definitions and assumptions from \autoref{sec:method} to represent a hypothetical PCT through SCM, and leverage \autoref{eq:itt}, \autoref{eq:at}, \autoref{eq:pp}, \autoref{eq:or}, \autoref{eq:rr}, and \autoref{eq:hr} on the dataset to find relevant treatment effects and outcome metrics.

\begin{figure}[htbp]
    \centering
    \includegraphics[width=0.5\linewidth]{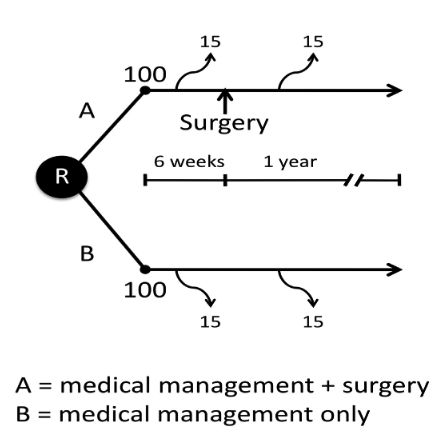}
    \caption{Hypothetical PCT in patients with cardiovascular disease. Intervention, A = medical management + surgery, vs. control, B = medical management only. Collected from McCoy et al. \citep{mccoy2017understanding}}
    ~\label{fig:data}
\end{figure}

For this purpose, we leverage a hypothetical pragmatic clinical trial, discussed in \citep{mccoy2017understanding} and presented in \autoref{fig:data}. In this PCT, an investigator conducted a study to evaluate whether the addition of surgery to a conventional medical therapy would benefit the patients (e.g., effective in controlling death in patients with cardiovascular disease). A total of two hundred (200) patients were enrolled, and half of them were allocated the new treatment protocol. The intervention treatment group received a combination of medical management and surgery, whereas the control group received only medical management. 

\paragraph{Assumption of Ground Truth}
With the usage of this dataset, we are also assuming the `ground truth' that the surgical intervention does not affect outcomes. Researchers are searching for this `ground truth'; one of the ways to do that is to conduct this hypothetical PCT.

\paragraph{Study Timeline Overview}
As shown in \autoref{fig:data}, after randomization, both arms of intervention contained a total of 100 patients. The medical management continues from randomization, and there is a timeline gap or waiting period of six (6) weeks from randomization to surgery. In treatment group A, 15 patients died before the six-week waiting period, and an additional 15 died between six weeks and 12 months. Similarly, 15 patients died before six weeks in treatment group B, and another 15 died between six weeks and 12 months.

\subsection{SCM for PCT}

\begin{figure}[htbp]
    \centering
    \includegraphics[width=0.6\linewidth]{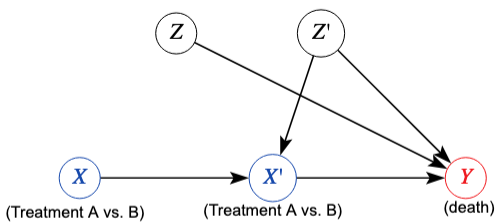}
    \caption{Graphical overview of SCM representation of the example PCT}
    ~\label{fig:hypo-pct-scm}
\end{figure}

Using definitions from \autoref{sec:method}, we represent the PCT through a SCM, as represented graphically in \autoref{fig:hypo-pct-scm}. $X$ is the treatment prescribed after randomization, where the population was divided equally between two treatment protocols. $X'$ is the treatment received, different from $X$ due to patients (count of 15) not going through surgery within six weeks. $Y$ is the outcome, death in a year for this trial. Although our graph shows $Z$ and $Z'$, we do not have any data on record on these two for this specific PCT.

\subsection{Outcome Analysis}

We reorganize the trial dataset to count patient outcomes for each value of $X$, $X'$, and $Y$. 

\begin{table}
    \centering
    \begin{tabular}{|l|l|l|l|}
    \hline
    \textbf{$X$} & \textbf{$X'$} & \textbf{$Y$} & \textbf{$Count$}\\
    \hline
    Treatment A & Treatment A & No death & 70\\ \cline{3-4}
    Treatment A & Treatment A & Death & 15\\  \cline{2-4}
    Treatment A & Treatment B & No death & 0\\  \cline{3-4}
    Treatment A & Treatment B & Death & 15\\  \cline{1-4}
    Treatment B & Treatment A & No death & 0\\  \cline{3-4}
    Treatment B & Treatment A & Death & 0\\  \cline{2-4}
    Treatment B & Treatment B & No death & 70\\  \cline{3-4}
    Treatment B & Treatment B & Death & 30\\
    \hline
    \end{tabular}
    \caption{Population distribution for different values of treatment prescribed $X$, treatment received $X'$ and outcome $Y$}
    ~\label{tab:hypo-pct-data}
\end{table}

Application of equations from \autoref{sec:method} are presented in \autoref{tab:hypo-pct-results}. The calculated results match with the results reported in \citep{mccoy2017understanding} and show that the equations discussed hold their originality, with the addition of SCM for a better understanding of the trial. Similar results can also be estimated through the equations provided from datasets used in other similar studies for PCT \citep{montori2001intention,johnston2016best}.

\begin{table}
    \centering
    \begin{tabular}{|l|l|l|l|}
    \hline
     & \textbf{ITT} & \textbf{AT} & \textbf{PP}\\
     \hline
    \textbf{RR} & $\frac{\frac{P(Y=0|X=1)}{P(Y=0|X=1)+P(Y=1|X=1)}}{\frac{P(Y=0|X=0)}{P(Y=0|X=0)+P(Y=1|X=0)}}$ & $\frac{\frac{P(Y=0|X'=1)}{P(Y=0|X'=1)+P(Y=1|X'=1)}}{\frac{P(Y=0|X'=0)}{P(Y=0|X'=0)+P(Y=1|X'=0)}}$ & $\frac{\frac{P(Y=0|X=1,X'=1)}{P(Y=0|X=1,X'=1)+P(Y=1|X=1,X'=1)}}{\frac{P(Y=0|X=0,X'=0)}{P(Y=0|X=0,X'=0)+P(Y=1|X=0,X'=0)}}$\\
    & $= \frac{P(Y=0|X=1)}{P(Y=0|X=0)}$ & $= \frac{P(Y=0|X'=1)}{P(Y=0|X'=0)}$ & $= \frac{P(Y=0|X=1,X'=1)}{P(Y=0|X=0,X'=0)}$\\
    & $= \frac{\frac{15+15}{70+15+0+15}}{\frac{0+30}{0+0+70+30}}$ & $= \frac{\frac{15+0}{70+15+0+0}}{\frac{15+30}{0+15+70+30}}$ & $= \frac{\frac{15}{70+15}}{\frac{30}{70+30}}$\\
    & $= \frac{0.3}{0.3}$ & $= \frac{0.18}{0.39}$ & $= \frac{0.18}{0.3}$\\
    & $= 1.00$ & $= 0.46$ & $= 0.60$\\ \cline{1-4}
    \textbf{OR} & $\frac{\frac{P(Y=0|X=1)}{P(Y=1|X=1)}}{\frac{P(Y=0|X=0)}{P(Y=1|X=0)}}$ & $\frac{\frac{P(Y=0|X'=1)}{P(Y=1|X'=1)}}{\frac{P(Y=0|X'=0)}{P(Y=1|X'=0)}}$ & $\frac{\frac{P(Y=0|X=1,X'=1)}{P(Y=1|X=1,X'=1)}}{\frac{P(Y=0|X=0,X'=0)}{P(Y=1|X=0,X'=0)}}$\\
    & $= \frac{\frac{0.3}{1-0.3}}{\frac{0.3}{1-0.3}}$ & $= \frac{\frac{0.18}{1-0.18}}{\frac{0.39}{1-0.39}}$ & $= \frac{\frac{0.18}{1-0.18}}{\frac{0.3}{1-0.3}}$\\
    % & $= \frac{\frac{0.3}{0.7}}{\frac{0.3}{0.7}}$ & $= \frac{0.18}{0.39}$ & $= \frac{0.18}{0.3}$\\
    & $= 1.00$ & $= 0.34$ & $= 0.51$\\
    \hline
    \end{tabular}
    \caption{Outcome metrics for the PCT}
    ~\label{tab:hypo-pct-results}
\end{table}

%=====================================================
\section{Discussion and Conclusion}

In this work, we have discussed the notion of leveraging structural causal models within causal inference to represent pragmatic clinical trials. Our proposition, along with relevant data analysis on the simulated PCT dataset, shows a prospective path of exploring PCTs for treatment effect estimations, counterfactual analyses, and transportability methods explorations.

% Uniqueness & Strengths
\paragraph{Strengths}
The essential contribution of this proposition is the notion of leveraging SCM for expressing PCTs. SCM and relevant causal inference methodologies have already been highly beneficial in estimating causal effects for different experimental and observational studies \citep{stovitz2019causal}. PCTs are highly meaningful for decision-makers as they are easier to conduct and convey treatment efficacy in a standard-setting. Since PCTs are more fluid in their nature than other experimental studies, the need to draw causal estimations from PCT is also higher than others. The uniqueness of this proposition is defined by the usage of $X$ and $X'$ representing treatments as two causally connected yet different variables.

% causal interpretation of guidelines
\paragraph{Causal Equivalent of Guidelines for PCT}
The four key design elements of PCT, by definition \citep{gamerman2019pragmatic}, are real-world population, real-world setting, appropriate comparison arm, and relevant outcome. Excluding only real-world settings, the SCM definition for PCT can utilize all the other elements. The concept also reflects and pairs perfectly with the guidelines provided by \citep{murray2019guidelines}.

% causal interpretation of equations
\paragraph{Causal Interpretation of Analysis Equations}
Given an OBS with $X$ as treatment, $Y$ as an outcome, and $Z$ as confounders, we easily find the conditional probability of outcome $Y$ given $X$ [$P(Y|X)$]. To find the equivalent causal effect, we either conduct a similar RCT with treatment randomized (aka de-confounded) and look at $P(Y|X)$ or simulate the RCT from the OBS using do-calculus ($P(Y|do(X))$). Resembling to that conversion of $P(Y|X)$ to $P(Y|do(X))$, we explore causal effects from the equations \autoref{eq:itt}, \autoref{eq:at}, and \autoref{eq:pp} by applying do-calculus on these. It results in:

\begin{equation}
~\label{eq:do-itt}
    P(Y | \mathit{do}(X)) = P(Y | X)
\end{equation}

\begin{equation}
~\label{eq:do-at}
    P(Y | \mathit{do}(X')) = \sum_{Z'} P(Y | X',Z') P(Z')
\end{equation}

\begin{equation}
~\label{eq:do-pp}
    P(Y | \mathit{do}(X=a), \mathit{do}(X'=a)) = \sum_{Z'} P(Y | X=a,X'=a,Z') P(Z')
\end{equation}

% however we look for the causal effect 
\autoref{eq:do-itt}, \autoref{eq:do-at}, and \autoref{eq:do-pp} provides two interesting insights to the notion proposed. 

\textbf{(1)} Since $X$ is randomized, \autoref{eq:do-at} is equal to its equivalent conditional probability equation. This estimation is the most standard (unbiased) estimation in providing treatment effect, which also aligns with \citep{montori2001intention}. Nevertheless, it still cannot minimize bias introduced by loss to follow-up, as $X'$ is not considered in this equation. \autoref{eq:do-at} does not use $X$ but uses $X'$, and is deconfounded by using $Z'$. The effect estimation is helpful, but the causal estimation requires a knowledge of measured confounders $Z'$, which is hard to find in the real world. This equation is also valuable since it shows the effect of non-adherence on the trial participants (through $X'$). \autoref{eq:do-pp} uses both $X$ and $X'$ in estimating the effect, by which it captures the essence of the population who strictly adhered to the protocol.

\textbf{(2)} Although $X$ and $X'$ represent treatment in different population percentages, they still fundamentally represent the same treatment for the study. While conducting a real-life PCT, with patients lost to follow-up, the ITT analysis results do not match with AT analysis results. Under normal conditions, $P(Y|X)$ and $P(Y|X')$ would never be equal. However, with do-calculus, it is expected that $P(Y|\mathit{do}(X))$ and $P(Y|\mathit{do}(X'))$ would be the same since they both indicate the causal effect of treatment on outcome. It raises the idea that, if we can identify a true set of confounders $Z'$ (that affects adherence), we can estimate the true causal effect of treatment on outcome from a PCT, and in those cases, \autoref{eq:do-itt}, \autoref{eq:do-at}, and \autoref{eq:do-pp} will all produce the same effect estimate.

% Weakness and Limitations
\paragraph{Limitations}
The prime challenge is defining the relevant causal structure for the SCMs representing the PCT. RCTs (and Obs) are frequently expressed through SCMs; however, that does not happen with PCT due to their pragmatic nature by definition. Researchers continuously explore ways to build causal structure through data and priors (background knowledge, peer-reviewed literature). Another critical challenge in this research is to find an appropriate set of confounders $Z'$. Confounding variables, in most cases, are not observed, measured, or even found. Finally, in PCTs, the treatment prescribed generally differs from the treatment received. Thus, adherence to the trial is vital, and causal effect estimation becomes complex when the information is unavailable or hard to determine. 

% Future Works
\paragraph{Future Works}
Our future work will include instrumental variable analysis \citep{bareinboim2012controlling}, by using treatment $X$ in \autoref{fig:scm-rct-pct} as an instrumental variable for the proposed causal graph. We will additionally explore time-series intervention with the definition proposed, in place of point intervention, by altering the SCM and related transportability equations.

% CURRENT VERSION
% \input{sections/latest_main_file}

\bibliography{sample}

% \appendix
% \section*{Appendix A.}

% Some more details about those methods, so we can actually reproduce
% them.  After the blind review period, you could link to a repository
% for the code also.  

\end{document}